\documentclass[10pt]{article}
\usepackage{float}
\usepackage[symbol]{footmisc} 
\usepackage[superscript,sort]{cite}
\usepackage{array}
\usepackage{bm}

\newcolumntype{x}[1]{%
>{\centering\hspace{0pt}}p{#1}}%
\usepackage[normalem]{ulem} 
\usepackage[colorlinks=true,linkcolor=black,citecolor=black,urlcolor=black,bookmarks=true,breaklinks=true]{hyperref}
\usepackage[usenames]{color}
\usepackage{amsmath,amssymb}

\usepackage{amsthm,amscd,amsxtra,amsfonts,amsmath,amssymb,multirow}
\usepackage{wrapfig}
\usepackage[footnotesize]{caption}
\usepackage{subcaption}
\usepackage[tiny,compact]{titlesec}
\usepackage{threeparttable} 
\usepackage{helvet}

\usepackage{booktabs}
\usepackage[table]{xcolor}
\usepackage{xcolor,colortbl}
\usepackage{tikz}
\usetikzlibrary{shapes,arrows}
\usepackage[T1]{fontenc}
\usepackage[linesnumbered,ruled]{algorithm2e} 
\titlespacing*{\section}{0pt}{*0}{*0}
\titlespacing*{\subsection}{0pt}{*0}{*0}
\titlespacing*{\subsubsection}{0pt}{*0}{*0} 
\titlespacing{\paragraph}{0pt}{*0}{*1}

\definecolor{MyPurple}{rgb}{1,0,1}

\newcommand{\beq}[1]{\begin{equation} \label{#1}}
\newcommand{\eeq}{\end{equation}}
\newcommand{\barray}{\begin{array}{ll}}
\newcommand{\earray}{\end{array}}

\newcommand\MutTestTOPPCCMedian{0.77}
\newcommand\MutTestTOPRMSEMedian{1.01}
\newcommand\MutTestALLPCCMedian{0.82} 
\newcommand\MutTestALLRMSEMedian{0.92}
\newcommand\MutTestGEOPCCMedian{0.76} 
\newcommand\MutTestGEORMSEMedian{1.03}
\newcommand\MutTestELCPCCMedian{0.76} 
\newcommand\MutTestELCRMSEMedian{1.02}
\newcommand\MutTestHLFPCCMedian{0.68} 
\newcommand\MutTestHLFRMSEMedian{1.14}
\newcommand\MutTestEVOSEQPCCMedian{0.61} 
\newcommand\MutTestEVOSEQRMSEMedian{1.26}
\newcommand\MutCVTOPPCCMedian{0.75}
\newcommand\MutCVTOPRMSEMedian{0.97}
\newcommand\MutCVALLPCCMedian{0.79}
\newcommand\MutCVALLRMSEMedian{0.90}
\newcommand\MutCVGEOPCCMedian{0.72}
\newcommand\MutCVGEORMSEMedian{1.03}
\newcommand\MutCVELCPCCMedian{0.72}
\newcommand\MutCVELCRMSEMedian{1.02}
\newcommand\MutCVHLFPCCMedian{0.66}
\newcommand\MutCVHLFRMSEMedian{1.11}
\newcommand\MutCVEVOSEQPCCMedian{0.62}
\newcommand\MutCVEVOSEQRMSEMedian{1.16}
\newcommand\MemALLPCCMedian{0.57}  
\newcommand\MemALLRMSEMedian{1.09}
\newcommand\MemTOPPCCMedian{0.54} 
\newcommand\MemTOPRMSEMedian{1.12}
\newcommand\MemGEOPCCMedian{0.48} 
\newcommand\MemGEORMSEMedian{1.17}
\newcommand\MemELCPCCMedian{0.53} 
\newcommand\MemELCRMSEMedian{1.14}
\newcommand\MemHLFPCCMedian{0.23} 
\newcommand\MemHLFRMSEMedian{1.41}
\newcommand\MemEVOSEQPCCMedian{0.38} 
\newcommand\MemEVOSEQRMSEMedian{1.26}

\begin{document}
\pagenumbering{roman}

\clearpage \pagebreak \setcounter{page}{1}
\renewcommand{\thepage}{{\arabic{page}}}

\title{Analysis and prediction of protein folding energy changes upon mutation by element specific persistent homology
}

\author{
Zixuan Cang$^1$,
 and
Guo-Wei Wei$^{1,2,3}$ \footnote{ Address correspondences  to Guo-Wei Wei. E-mail:wei@math.msu.edu}\\
$^1$ Department of Mathematics \\
Michigan State University, MI 48824, USA\\
$^2$  Department of Biochemistry and Molecular Biology\\
Michigan State University, MI 48824, USA \\
$^3$ Department of Electrical and Computer Engineering \\
Michigan State University, MI 48824, USA \\
}

\date{}
\maketitle
\maketitle
\abstract{
\noindent
\textbf{Motivation:}
Site directed mutagenesis is widely used to understand the structure and function of biomolecules. Computational prediction of protein mutation impacts offers a fast, economical and potentially accurate alternative to laboratory mutagenesis.  Most existing methods rely on geometric descriptions, this work introduces a topology based approach to provide  an entirely new representation of protein mutation impacts  that could not be  obtained from conventional techniques. \\
\textbf{Results:} Topology based mutation predictor (T-MP) is introduced to dramatically reduce the geometric complexity and number of  degrees of freedom  of proteins, while element specific persistent homology is proposed to retain essential biological information. The present approach is found to outperform other existing methods in  globular protein mutation impact predictions. A Pearson correlation coefficient of 0.82 with an RMSE of 0.92 kcal/mol is obtained on a test set of 350 mutation samples. For the prediction of membrane protein stability changes upon mutation, the proposed topological approach has a 84\% higher Pearson correlation coefficient than the current state-of-the-art empirical methods, achieving a Pearson correlation of 0.57 and an RMSE of 1.09 kcal/mol in a 5-fold cross validation on a set of 223 membrane protein mutation samples. \\

\section{Introduction}
Mutagenesis, as a basic biological process that changes the genetic information of organisms, serves as a primary  source for many kinds of cancer and heritable diseases, as well as a driving force for natural evolution \cite{PYue:2005,ZZhang:2012, Kucukkal:2015}. For example, more than 60 human hereditary diseases are directly related to mutagenesis in proteases and their natural inhibitors   \cite{Puente:2003}. Additionally, mutagenesis  often leads to drug resistance  \cite{Martinez:2000}.  Mutation, as  a  result of  mutagenesis,  can either occur  spontaneously in nature  or  be caused by the exposure to a large dose of mutagens in living organisms. In laboratories, site directed mutagenesis analysis is a vital experimental  procedure for exploring protein functional changes in enzymatic catalyzing, structural supporting, ligand binding, and signaling \cite{Fersht:2078}. Nonetheless, site directed mutagenesis analysis is both time-consuming and expensive. Additionally,  site directed mutagenesis measurements for one specific mutation obtained from different approaches may vary dramatically, particularly for membrane protein mutations. 

Computational prediction of protein mutation impacts is an important alternative to experimental mutagenesis analysis for the systematical  exploration of  protein structural instabilities, functions, disease connections, and organism evolution pathways \cite{Guerois:2002}. A major advantage of these approaches is that they provide an economical, fast, and potentially accurate alternative to site directed mutagenesis experiments. Many  state-of-the-art  methods have been developed in the past decade,  
  including I-Mutant \cite{Capriotti:2005}, PoPMuSiC \cite{Dehouck:2009}, knowledge-modified MM/PBSA approach  \cite{Getov:2016},  Rosetta (high) protocols \cite{Kellogg:2011},    FoldX (3.0, beta 6.1) \cite{Guerois:2002},   SDM \cite{Worth:2011},  DUET \cite{Pires:2014b},  PPSC (Prediction of Protein Stability, version 1.0) with the 8 (M8) and 47 (M47) feature sets \cite{YYang:2013},  PROVEAN \cite{YChoi:2012}, ELASPIC  \cite{Berliner:2014}, STRUM \cite{LJQuan:2016}, and EASE-MM \cite{Folkman:2016}.  
 In general, computational approaches can be classified into three major classes. Among them, physics based methods typically  make use of molecular mechanics (MM), quantum mechanics (QM), or multiscale implicit solvent models and QM/MM approaches. These approaches elucidate the fundamental of physics and offer physical insights to mutagenesis. Empirical models are another class of methods that utilize  empirical functions and potential terms to describe  mutation impacts. Model parameters are fit with a given set of experimental data and the resulting model is used to predict new mutation induced folding free energy changes. The last class of approaches is knowledge based methods that invoke modern machine learning techniques to uncover hidden relationships between protein stability and protein structure as well as sequence. A major advantage of knowledge based mutation predictors is their ability to handle increasingly large and diverse mutation data sets. However, the performance of these approaches highly depends on the training sets and their results usually can not be easily interpreted in physical terms. 

A common challenge for all existing mutation impact prediction models is in achieving  accurate and reliable predictions  of membrane protein stability changes upon mutation. As recently noted by  Kroncke {\it et al}, currently there is no reliable method for the prediction of membrane protein mutation impacts \cite{Kroncke:2016}. The membrane protein mutation data set studied by these authors has fewer than 250 data entries, which is too few for most knowledge based methods, and involves 7 membrane protein families, which is too diverse for typical physics based methods. 
  
\begin{figure}[ht]
\begin{center}
\includegraphics[keepaspectratio,width=0.4\columnwidth]{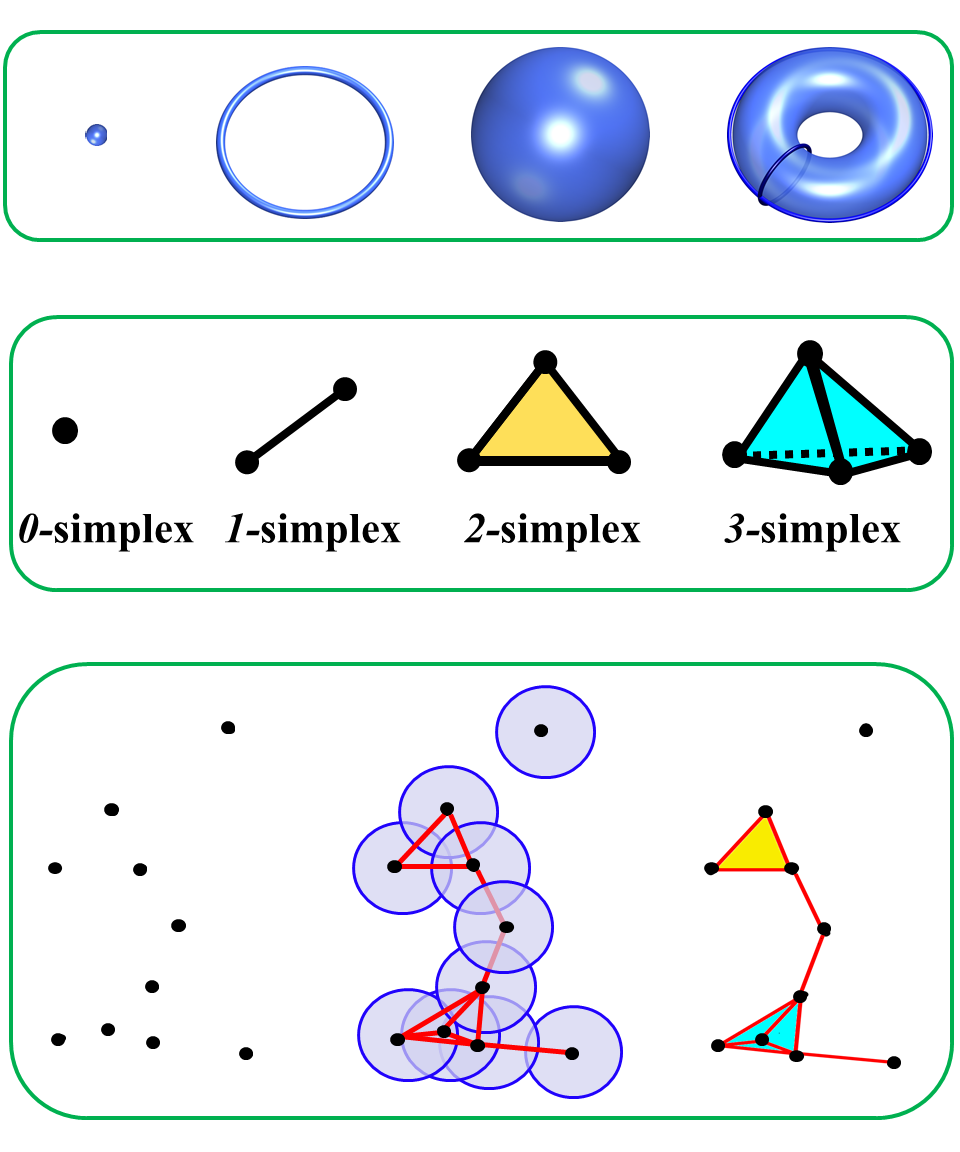}
\caption{An illustration of topological invariants (Top row), basic simplexes (Middle row) and simplicial complex construction in a given radius of filtration (Bottom row). 
Top row: a point, a circle, an empty sphere and a torus are displayed from left to right. Betti-0, Betti-1 and Betti-2  numbers for point are, respectively, 1,0 and 0, for the circle 0,1 and 0, for the empty sphere 0,0 and 1, and for the torus 1,2 and 1.         
Two auxiliary rings are added to the torus explain Betti-1$=2$.  
Middle row: Four typical simplexes are illustrated. 
Bottom row:  Illustration of a set of ten points (left chart) at a given filtration radius (middle chart) and the corresponding simplicial complexes (right chart), where there are  one 0-simplex, three 1-simplexes, one 2-simplex and one 3-simplex.  
      }
\label{fig:invariants}
\end{center}
\end{figure}

A key feature of all existing structure based mutation impact predictors is that they either fully or partially rely on direct geometric descriptions which rest in  excessively  high dimensional spaces resulting in large number of degrees of freedom. In practice, the geometry can easily be over simplified. Mathematically, topology, in contrast to geometry, concerns the connectivity of different components in a space \cite{kaczynski:mischaikow:mrozek:04}, and  offers the ultimate level of abstraction of data.  However, conventional topology incurs too much reduction of geometric information to be practically useful in biomolecular analysis. Persistent homology, a new branch of algebraic topology, retains partial geometric information in topological description, and thus bridges the gap between geometry and topology  \cite{Edelsbrunner:2002,Zomorodian:2005}. It has been applied to biomolecular characterization, identification and analysis  \cite{YaoY:2009, KLXia:2014c, Mate:2014, KLXia:2015c, BaoWang:2016a}. However, conventional  persistent homology makes no distinction of different atoms in a biomolecule, which results in a heavy loss of  biological information and limits its  performance in protein classification   \cite{ZXCang:2015}. \\
In the present work, we introduce element specific persistent homology (ESPH), interactive persistent homology and binned barcode representation  to retain essential biological information in the topological simplification of biological complexity. We further integrate ESPH and machine learning  to analyze and predict protein mutation impacts. The essential idea of our topological mutation predictor (T-MP) is to use ESPH to  transform the biomolecular data in the high-dimensional space with full biological complexity to a space of fewer dimensions and simplified biological complexity, and to use machine learning to deal with  massive and diverse data sets. A distinct feature  of the present T-MP is that the prediction results can be analyzed and interpreted in physical terms to shed light on the molecular mechanism of protein folding energy changes upon mutation. Additionally, the mathematical model for different types of mutations can be adaptively optimized according to the performance analysis of ESPH features. We demonstrate that the performance of proposed T-MP matches or excesses that of other existing methods. \\   

\section{Methods}
\subsection{Persistent homology characterization of proteins}
Unlike physics based models which describe protein folding in terms of covalent bonds, hydrogen bonds, electrostatic and  van der Waals interactions, the natural language of persistent homology  is topological invariants, i.e., the intrinsic features of the underlying topological space. More specifically,  independent components, rings and cavities are topological invariants in a given data set and their numbers are called Betti-0, Betti-1 and Betti-2, respectively, as shown in the top row of Fig. \ref{fig:invariants}. Loosely speaking,    simplicial complexes are generated from discrete data points according to a specific rule such as Vietoris-Rips   complex, C$\check{e}$ch complex, or alpha complex.  Specifically, a 0-simplex is a vertex, a 1-simplex is an edge, a 2-simplex is a triangle, and a 3-simplex represents a tetrahedron, see the middle row of Fig. \ref{fig:invariants}. Algebraic groups built on these simplicial complexes are used in simplicial homology to  practically compute Betti numbers of various dimensions. Furthermore, persistent homology creates a series of homologies through a filtration process, in which the connectivity of a given data set is systematically reset according to a scale parameter, such as  an ever-increasing  radius of every atom in a protein, see the bottom row of  Fig. \ref{fig:invariants}.  As a result, the birth, death, and  persistence of  topological invariants over the filtration give rise to the barcode representation of a given data set  \cite{Ghrist:2008}.  When persistent homology is used to analyze  three dimensional (3D) protein structures,  one-dimensional (1D) persistent homology barcodes are obtained as topological fingerprints (TFs) \cite{YaoY:2009, KLXia:2014c, Mate:2014, ZXCang:2015}.  

As an illustration, we consider  the persistent homology analysis of a wild type protein (PDB:1ey0) and its mutant. The mutation (G88W) occurred at residue 88 from Gly to Typ is shown at Fig.  \ref{fig:MutationBarcodes}{\bf a} and {\bf b}. In this case, a small residue (Gly) is replaced by a large one (Typ).  We carry out persistent homology analysis of a set of heavy atoms within 6\AA~ from the mutation site. Persistent homology barcodes of the wild type and the mutant are respectively given  in Fig.  \ref{fig:MutationBarcodes} {\bf c} and {\bf d}, where the three panels from top to bottom are for  Betti-0, Betti-1, and Betti-2, respectively. Since the set of atoms included in the wild type and the mutant is the same except for that in the mutation site, the obvious difference in  persistent homology barcodes is induced by the mutation. The increase of residue size results in tighter parttern of Betti-0 bars where there are fewer relatively long bars and more Betti-1 and Betti-2 bars in a shorter distance scale are observed.

\begin{figure}[ht]
\begin{center}
\includegraphics[keepaspectratio,width=0.5\columnwidth]{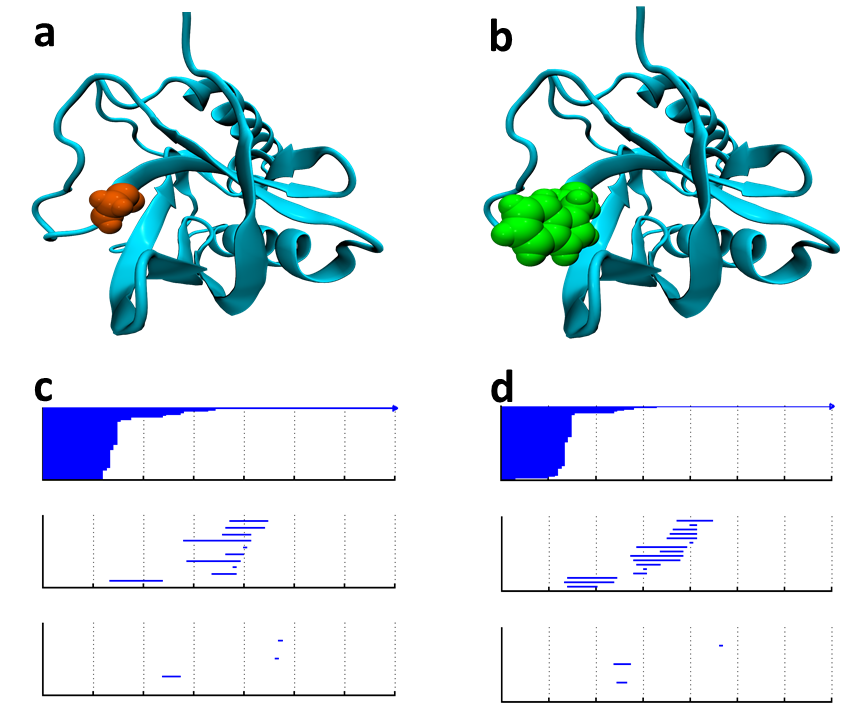}
\caption{An illustration of persistent homology   barcode changes from wild type  to  mutant proteins. 
\textbf{a} The wild type protein (PDB:1ey0) with residue 88 as Gly. 
\textbf{b} The mutant with residue 88 as Typ. 
\textbf{c} Wild type protein barcodes for  heavy atoms  within 6 \AA~ of the mutation site.   Three panels  from top to bottom are Betti-0, Betti-1, and Betti-2 barcodes, respectively. The horizontal axis is the filtration radius (\AA).  
\textbf{d} Mutant protein barcodes obtained similarly as those for  the wild type.
}
\label{fig:MutationBarcodes}
\end{center}
\end{figure}

Nonetheless, the above topological representation of proteins does not contain sufficient  biological information, such as bond length distribution of a given type of atoms, hydrogen bonds, hydrophobic and hydrophilic effects, to offer an accurate model for protein mutation impact predictions.  To characterize chemical and biological properties of  biomolecules,  we introduce {\it element specific persistent homology} (ESPH). Instead of labeling  every  atom as in many physics based methods, we distinguish different element types of  biomolecules in constructing persistent homology barcodes. For proteins, commonly occurring   element types include  ${\rm C, N, O, S}$ and ${\rm H}$. Among them,  hydrogen atoms are often absent from  PDB data and sulfur atoms are too few to be statistically significant in most proteins. Therefore, we focus on the ESPH of ${\rm C, N}$ and ${\rm O}$ elements in protein characterization. 

\begin{figure}[ht]
\begin{center}
\includegraphics[keepaspectratio,width=0.5\columnwidth]{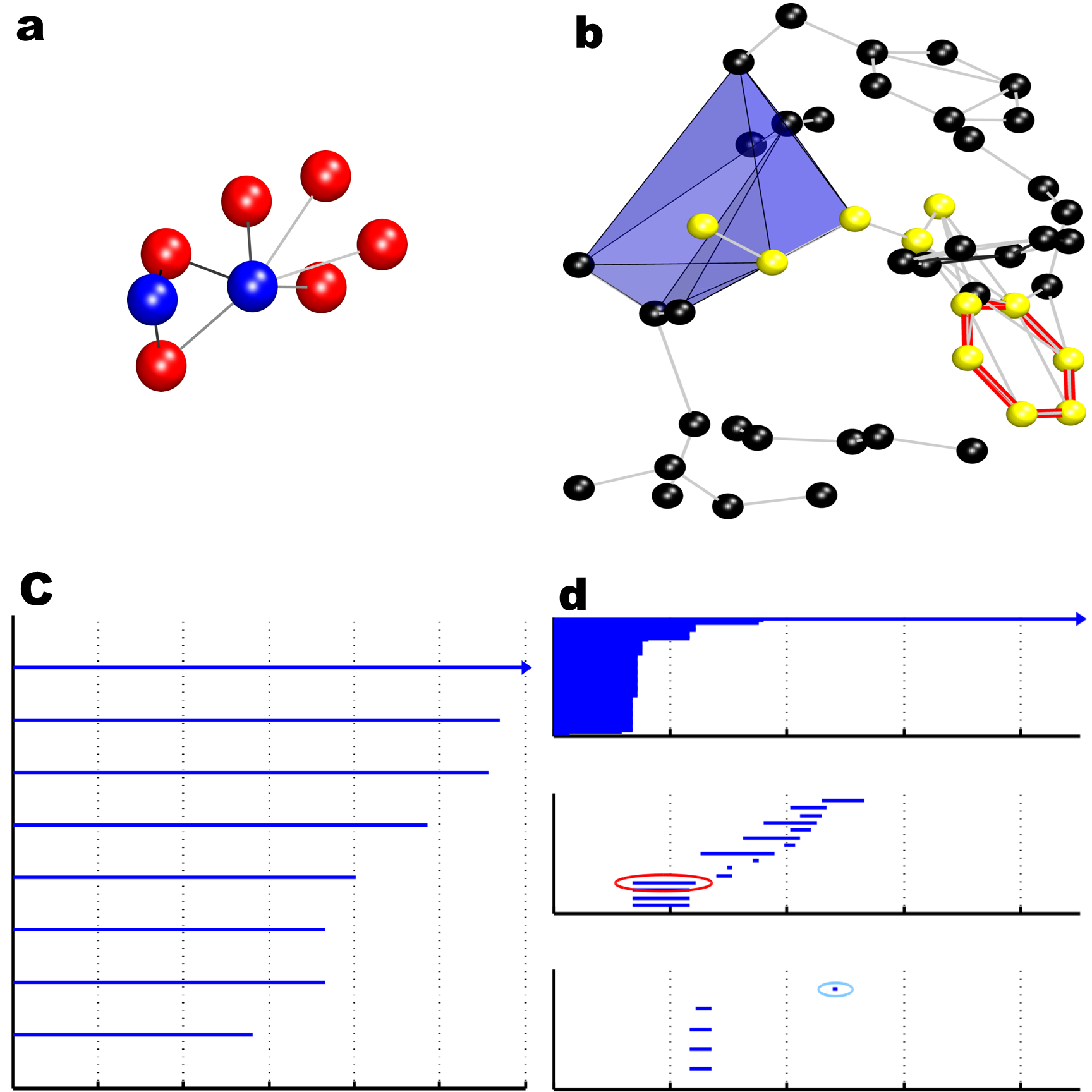}
\caption{An illustration of element specific persistent homology (ESPH) indicating the hydrophilic network (Left) and hydrophobic network (Right) at a mutation site.   
{\bf a}: Hydrophilic network showing the connectivity  between nitrogen atoms of the mutation residue  (blue) and oxygen atoms of the rest of the protein (red). 
{\bf b}: Hydrophobic network showing the connectivity  between carbon atoms of the rest of the protein (black) and of the mutation residue (yellow). 
 Red circles label a hexagon ring  and blue filling indicating a cavity. 
{\bf c}:  The ESPH Betti-0 barcodes of the hydrophilic network in {\bf a}.  Betti-0 barcodes show not only the number and strength of 
of  hydrogen bonds, but also the hydrophilic environment. Specifically, the shortest four bars can be directly interpreted as conventional hydrogen bonds, while other bars contributing the degree of hydrophilicity at the mutation site. 
{\bf d}:  The ESPH Betti-0, Betti-1 and Betti-2 barcodes  of  the hydrophobic  network in {\bf b}. The bar in the red circle is due to the hexagon ring in {\bf b} and the bar in the blue circle is due to the cavity in {\bf b}.  
}
\label{Figure3}
\end{center}
\end{figure}

\subsection{Topological descriptors}
The most important issue in protein mutation impact analysis is the interactions between the mutation site and the rest of the protein. To describe these interactions, we propose {\it interactive persistent homology} adopting the distance function $DI(A_i, A_j)$ describing the distance between two atoms $A_i$ and $A_j$ defined as 
\begin{equation}
DI(A_i, A_j) = \begin{cases}
\infty, \, if \, {\rm Loc}(A_i)={\rm Loc}(A_j),  \\
DE(A_i, A_j), \, otherwise,
\end{cases}
\end{equation}
where $DE(\cdot, \cdot)$ is the Euclidean distance between the two atoms and ${\rm Loc}(\cdot)$ denotes the location of an atom which is either in a mutation site or in the rest of the protein. In the persistent homology computation, Vietoris-Rips complex (VC) and alpha complex (AC) are used for characterizing first order interactions and higher order patterns respectively. To characterize interactions of different kinds, we construct persistent homology barcodes on the atom sets by selecting one certain type of atoms in mutation site and one other certain type of atoms in the rest of the protein. We denote the set of bar codes from one persistent homology computation as $V^{p,d,b}_{\gamma,\alpha,\beta}$ where $p \in\{{\rm VC,AC}\}$ is the complex used, $d\in\{DI,DE\}$ is the distance function, $b\in\{0,1,2\}$ represents the topological dimensions, $\alpha\in\{\rm{C,N,O}\}$ is the element type selected in the rest of the protein, and $\beta\in\{\rm{C,N,O}\}$ is the element type selected in the mutation site. $\gamma\in\{{\rm M,W}\}$ denotes whether the mutant protein or the wild type protein is used for the calculation. The proposed approach ends up with a total of 54 sets of persistent homology bar codes ($V^{{\rm VC},DI,0}_{\gamma, \alpha, \beta}$, where $\alpha,\beta=\rm{C,N,O}; \gamma = {\rm M,W}$ and $V^{{\rm AC},DE,b}_{\gamma,\alpha,\beta}$, where $\alpha,\beta=\rm{C,N,O}; \gamma={\rm M,W}; b=1,2$). These barcodes are capable of revealing the molecular mechanism of protein stability.  For example, interactive ESPH barcodes generated from  carbon atoms are associated with hydrophobic interaction networks in proteins. Similarly,  interactive ESPH barcodes between  nitrogen and oxygen atoms correlate to hydrophilic interactions and/or hydrogen bonds  as shown in Fig. \ref{Figure3}.
Interactive ESPH barcodes are also able to reveal other bond information; notwithstanding, they can not always be  interpreted as covalent bond,   hydrogen bonds, or van der Waals bonds in general.  In fact, interactive ESPH barcodes provide an entirely new representation of molecular interactions. 

Features are extracted from the groups of persistent homology barcodes. For the 18 groups of Betti-0 ESPH barcodes, though they cannot be literally interpreted as bond lengths, they can be used to effectively characterize biomolecular interactions. Interatomic distance is a crucial parameter for interaction strength. One can classify hydrogen bonds with donor-acceptor distances of 2.2-2.5\AA~ as strong and mostly covalent, 2.5-3.2\AA~ as moderate and mostly electrostatic, and 3.2-4.0\AA~  as weak and electrostatic  \cite{Jeffrey}. Their corresponding energies are about 40-14, 15-4, and less than 4 kcal/mol, respectively \cite{Jeffrey}. To  differentiate the interaction distances between various element types, we propose {\it binned barcode representation} (BBR)
by dividing interactive ESPH  barcodes into a number of equally spaced  bins, namely  $[0, 0.5], (0.5, 1], \cdots, (5.5, 6]$\AA. The death value of bars are counted in each bin resulting in 12*18 features. Such representation enables us to precisely characterize hydrogen bond, van der Waals, electrostatic,  hydrophilic and hydrophobic interactions. For the higher order Betti numbers, the emphasize is given on patterns of both short and long distance scales. Seven features are computed for each group of barcodes for Betti-1 or Betti-2 which are summation, maximum, and average bar length as well as maximum and minimum birth and death values resulting in 7*36 features. To contrast the interactive ESPH barcodes of wild type protein and mutant, we also take the differences between the features described above, which gives rise to a total of 702 features.

\subsection{Auxiliary descriptors}\label{sec:AuxFeature}
While the topological descriptors give a through examination of the atomic arrangements and interactions, some other crucial properties are not explicitly characterized. Additionally, due to the diverse quality of the structures examined, some higher level descriptors such as residue level descriptors can enhance the robustness of the model. Therefore, we include some auxiliary descriptors from the aspect of geometry, electrostatics, amino acid types composition, and amino acid sequence. The geometric descriptors contain surface area and van der Waals interaction. The electrostatics descriptors are consisted of atomic partial charge, Coulomb interaction, and atomic electrostatic solvation energy. The high level descriptors include neighborhood amino acid composition and predicted pKa shifts. The sequence descriptors describe the secondary structure and residue conservation score collected from Position-specific scoring matrix (PSSM). Details can be found in supplementary material.

\subsection{Gradient boosting trees regressor}
\begin{figure}[ht]
\begin{center}
\includegraphics[keepaspectratio,width=0.8\columnwidth]{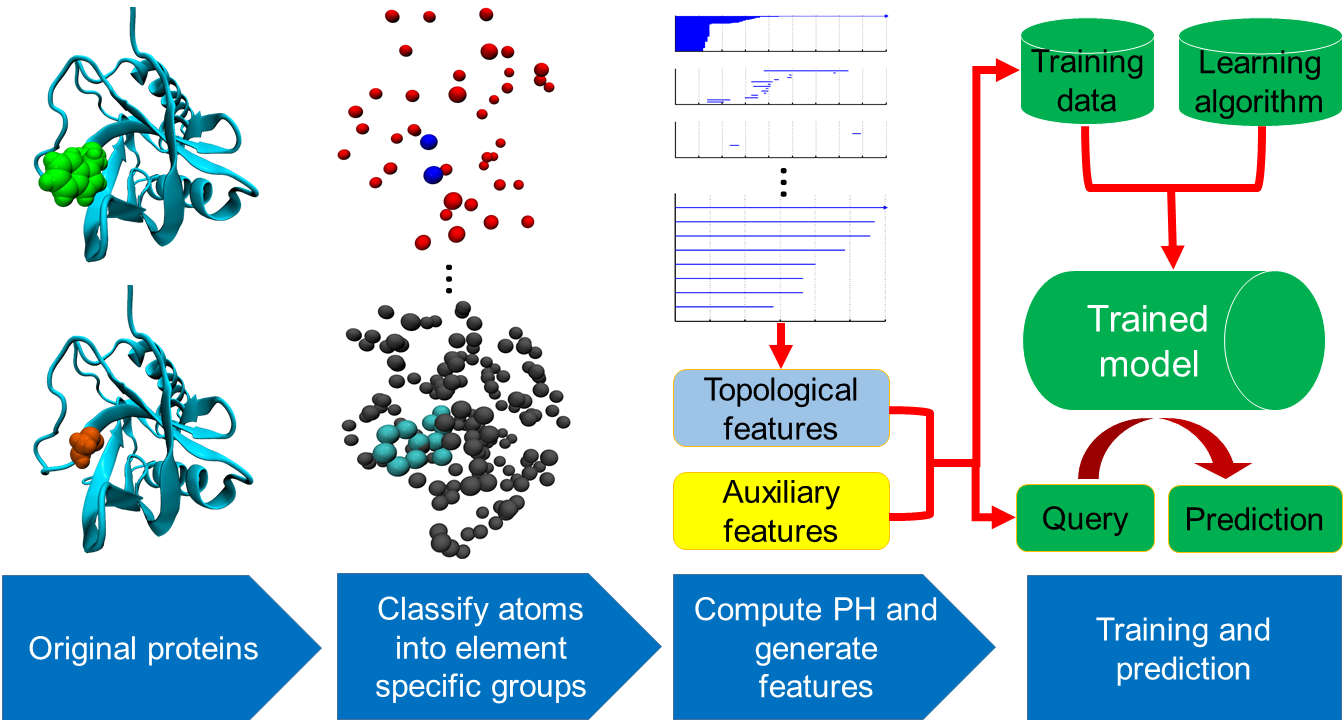}
\caption{Flowchart for topology based predictions of protein folding stability changes upon mutation.}
\label{fig:flowchart}
\end{center} 
\end{figure}
The topological features and the auxiliary features are ideally suited for being used as machine learning features to predict protein stability changes upon mutation.  Figure \ref{fig:flowchart} shows a schematic illustration of our T-MP. We have examined a number of machine learning algorithms, including  decision tree learning, random forest, and  gradient boosted regression trees (GBRTs) \cite{GBDT:2001}, in our study and found very similar results from these algorithms for the above binned interactive ESPH barcodes.  For example, GBRTs are able to integrate weak learners  to form a strong predictor. GBRTs uncover the coupling or nonlinear dependence among highly interactive topological features  by choosing an appropriate maximum tree depth. Additionally, GBRTs bypass the  normalization of  the topological feature vectors, and thus allow mixed attributes of different topological measures and physical units. Finally, GBRTs can effectively avoid overfitting by lowering the learning rate and carrying out subsampling, which is important in dealing with small training data sets, such as data sets for membrane protein mutations. 

\subsection{Dataset preparation}

The PDB files of  wild type proteins are downloaded from Protein Data Bank (PDB) \cite{Berman:2000}. The chains that contain  the mutation site are extracted and saved using VMD software package \cite{VMD}. The missing heavy atoms and hydrogen atoms are added to the structure using the Profix utility of Jackal software package \cite{ZXiang:2001}. The mutant protein structure is obtained using Scap utility from Jackal software package \cite{ZXiang:2001} by replacing the side chain of the mutation site with \emph{min} option being set to 4 where additional conformers obtained by perturbing the conformers in the rotamer library are explored.  Mutation energy changes are obtained from the ProTherm database \cite{Bava:2004}.

\section{Results}

\subsection{General performance}
\begin{table}[ht]
\centering
\rowcolors{2}{gray!25}{white}
\begin{tabular}{lcccccc}
\toprule
\rowcolor{gray!75}
Method & \multicolumn{3}{c}{S350} & \multicolumn{3}{c}{S2648} \\
\rowcolor{gray!75}
 & $n^d$ & $R_P$ & RMSE  & $n^d$ & $R_P^e$ & RMSE$^f$ \\
\midrule
 T-MP-2 & 350 & \MutTestALLPCCMedian & \MutTestALLRMSEMedian & 2648 & \MutCVALLPCCMedian & \MutCVALLRMSEMedian\\
 STRUM$^b$        & 350 & 0.79 & 0.98 & 2647 & 0.77 & 0.94 \\
 T-MP-1 & 350 & \MutTestTOPPCCMedian & \MutTestTOPRMSEMedian & 2648 & \MutCVTOPPCCMedian & \MutCVTOPRMSEMedian\\
 mCSM$^{b,c}$     & 350 & 0.73 & 1.08 & 2643 & 0.69 & 1.07 \\
 INPS$^{b,c}$     & 350 & 0.68 & 1.25 & 2648 & 0.56 & 1.26 \\
 PoPMuSiC 2.0$^b$ & 350 & 0.67 & 1.16 & 2647 & 0.61 & 1.17 \\
 PoPMuSiC 1.0$^a$ & 350 & 0.62 & 1.23 & - & - & - \\
 I-Mutant 3.0$^b$ & 338 & 0.53 & 1.35 & 2636 & 0.60 & 1.19 \\
 Dmutant$^a$      & 350 & 0.48 & 1.38 & - & - & - \\ 
 Automute$^a$     & 315 & 0.46 & 1.42 & - & - & - \\
 CUPSAT$^a$       & 346 & 0.37 & 1.46 & - & - & - \\
 Eris$^a$         & 334 & 0.35 & 1.49 & - & - &  - \\
 I-Mutant 2.0$^a$ & 346 & 0.29 & 1.50 & - & - & - \\
\bottomrule
\end{tabular}
\vspace{3mm}
\caption{Comparison of  Pearson correlation coefficients ($R_P$) and RMSEs (kcal/mol) of various methods on the prediction task of  the S350 set and 5-fold cross validation of the S2648. T-MP-1 is our topological based mutation predictor that solely utilizes structural information. T-MP-2 is our model that complements T-MP-1 with additional electrostatic, evolutionary and sequence information. The T-MP methods are tested with 50 repeated experiments and the medians are reported. $^a$ Data directly obtained from \cite{Worth:2011}. $^b$ Data obtained from \cite{LJQuan:2016}. $^c$ The results reported in the publications are listed in the table, however, according to \cite{LJQuan:2016},  the data from the online server has $R_p$(RMSE) of 0.59(1.28) and 0.70(1.13) for INPS and mCSM respectively in the task of S350 set. $^d$ Number of samples successfully processed.  }
\label{tab:MutationPerformance}
\end{table}

To demonstrate the power of the proposed T-MP for protein mutation impact predictions, we consider a data set of 2648 mutation 
instances in 131 proteins, called S2648 data set \cite{Dehouck:2009}. Additionally, a subset of the S2648 data set involving 67 proteins, named S350 set, is used as  a test set.  All thermodynamic data entries for these two  data sets are obtained from the ProTherm database \cite{Bava:2004}.
The present study involves two tasks \cite{LJQuan:2016},  namely, five-fold cross validations over the  S2648 set and the prediction of the test set, S350, using the rest of the S2648 set (i.e., 2298 instances) as the training data.

\begin{table}[h]
\centering
\rowcolors{2}{gray!25}{white}
\begin{tabular}{lcccccc}
\toprule
\rowcolor{gray!75}
Features & \multicolumn{2}{c}{S350} & \multicolumn{2}{c}{S2648} & \multicolumn{2}{c}{M223}\\
\rowcolor{gray!75}
  & $R_P$ & RMSE  & $R_P$ & RMSE & $R_P$ & RMSE \\
\midrule
 T-MP-2  & \MutTestALLPCCMedian & \MutTestALLRMSEMedian  & \MutCVALLPCCMedian & \MutCVALLRMSEMedian & \MemALLPCCMedian & \MemALLRMSEMedian\\
T-MP-1 & \MutTestTOPPCCMedian & \MutTestTOPRMSEMedian &  \MutCVTOPPCCMedian & \MutCVTOPRMSEMedian & \MemTOPPCCMedian & \MemTOPRMSEMedian\\
 E-MP & \MutTestELCPCCMedian & \MutTestELCRMSEMedian  & \MutCVELCPCCMedian & \MutCVELCRMSEMedian & \MemELCPCCMedian & \MemELCRMSEMedian\\
 G-MP & \MutTestGEOPCCMedian & \MutTestGEORMSEMedian  & \MutCVGEOPCCMedian & \MutCVGEORMSEMedian & \MemGEOPCCMedian & \MemGEORMSEMedian\\
 S-MP & \MutTestEVOSEQPCCMedian & \MutTestEVOSEQRMSEMedian  & \MutCVEVOSEQPCCMedian & \MutCVEVOSEQRMSEMedian & \MemEVOSEQPCCMedian & \MemEVOSEQRMSEMedian\\
 H-MP & \MutTestHLFPCCMedian & \MutTestHLFRMSEMedian  & \MutCVHLFPCCMedian & \MutCVHLFRMSEMedian & \MemHLFPCCMedian & \MemHLFRMSEMedian\\
\bottomrule
\end{tabular}
\vspace{3mm}
\caption{Pearson correlation coefficients and RMSEs in the unit of kcal/mol of auxiliary features for three tasks.  The medians of 50 repeated runs are reported. Here S350 is a test and its predictions are generated with a model trained with the training set S2648 excluding set S350. Results for S2648 are obtained from 5-fold cross validation. Similarly results for M223 are obtained from 5-fold validation.  Here G-MP, E-MP, H-MP and S-MP denote 
mutation predictors derived respectively from 
geometric features, electrostatic features, high level features, and 
sequence features described in Section \ref{sec:AuxFeature} and Supplementary data.
 }
\label{tab:AuxMutationPerformance}
\end{table}

A comparison of the performances of various  methods is summarized in Table \ref{tab:MutationPerformance}. Pearson correlations coefficient ($R_P$) and RMSE for  test set S350, and five-fold cross validations for training set S2648, are given for various methods, including ours.   The proposed topology based mutation predictor, labeled as T-MP-1, significantly outperforms other existing methods, except for  STRUM \cite{LJQuan:2016}. STRUM is constructed by using various descriptors including geometric, evolutionary and sequence information and its $R_P$ and RMSE are 0.79 and 0.98 kcal/mol, respectively for test set S350, and 0.77 and 0.94 kcal/mol, respectively for cross validation of S2648 set \cite{LJQuan:2016}.  STRUM's excellent performance motivates us to consider auxiliary features. To this end, we add features generated from geometric, electrostatic and sequence information (see Supplementary data) to our T-MP to construct T-MP-2.  As shown in  Table \ref{tab:MutationPerformance}, T-MP-2 has the best performance among all methods with $R_P$ and RMSE being 0.82 and 0.92 kcal/mol, respectively for test set S350, and 0.79 and 0.90 kcal/mol, respectively for cross validation of S2648 set.  A comparison between T-MP-1 and T-MP-2 indicates that  geometric, electrostatic  and sequence  features give rise to approximately 5\% improvement over the original topological prediction, indicating the importance of geometric, electrostatic   and sequence information to mutation predictions. 
However, as shown in Table \ref{tab:AuxMutationPerformance}, none of these features has more predictive power than the present topological descriptor.

\begin{figure}[ht]
\centering
	 \includegraphics[keepaspectratio,width=0.7\columnwidth]{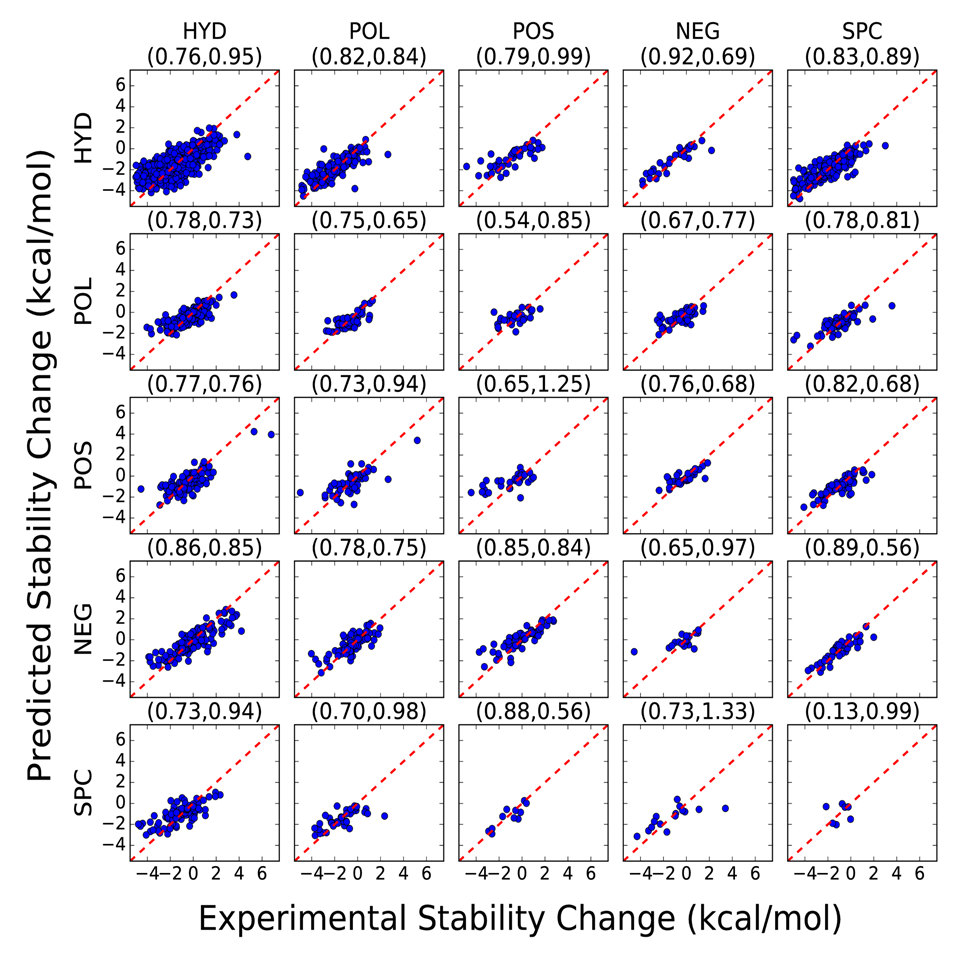}  
	\caption{Correlations between experimental mutation impacts and predicted stability changes (kcal/mol) upon mutation for 25 subsets of the S2648 data set. All predictions are obtained from 5-fold cross validations.  	For each subfigure, two numbers in brackets are  Pearson correlation coefficients and RMSEs (kcal/mol), respectively.  The vertical residue label and    horizontal residue label are respectively for the wild type and the mutant such that the second subfigure in the first row denotes a group of mutations from  hydrophobic residues (HYDs) to polar residues (POLs). The median is taken among 50 repeated experiments.   
		}  
	\label{fig:Correlations}
\end{figure}
\subsection{Performance in various mutation situations}\label{sec:PVMS}
Figure \ref{fig:Correlations} depicts detailed correlations between experimental mutation impacts and T-MP-2 predictions for 25 subsets of 2648 mutations from the cross validation process on the S2648 set. To this end, we adopt a standard classification that categorizes amino acid residues into hydrophobic (HYD), polar (POL), positively charged (POS), negatively charged (NEG) and special case (SPC) types. First, the majority of mutations lead to more unstable structures (i.e., negative free energy changes), as they should be.  However, two mutations from POS to HYD and one mutation from POS to POL lead to unusual  stabilizing effects. Moreover, the most accurate prediction in terms of RMSE was for a set of negatively charged residues  being mutated to special case ones. Not surprisingly, the worst performance is observed for mutations where little geometric rearrangements happen such as when a negatively charged residue is mutated to another residue of the same type. This performance analysis provides a guidance of how confident the prediction model is in different mutation situations.

\subsection{Performance of features of various element combinations}

	\begin{figure}[ht]
	\centering
	 \includegraphics[keepaspectratio,width=0.8\columnwidth]{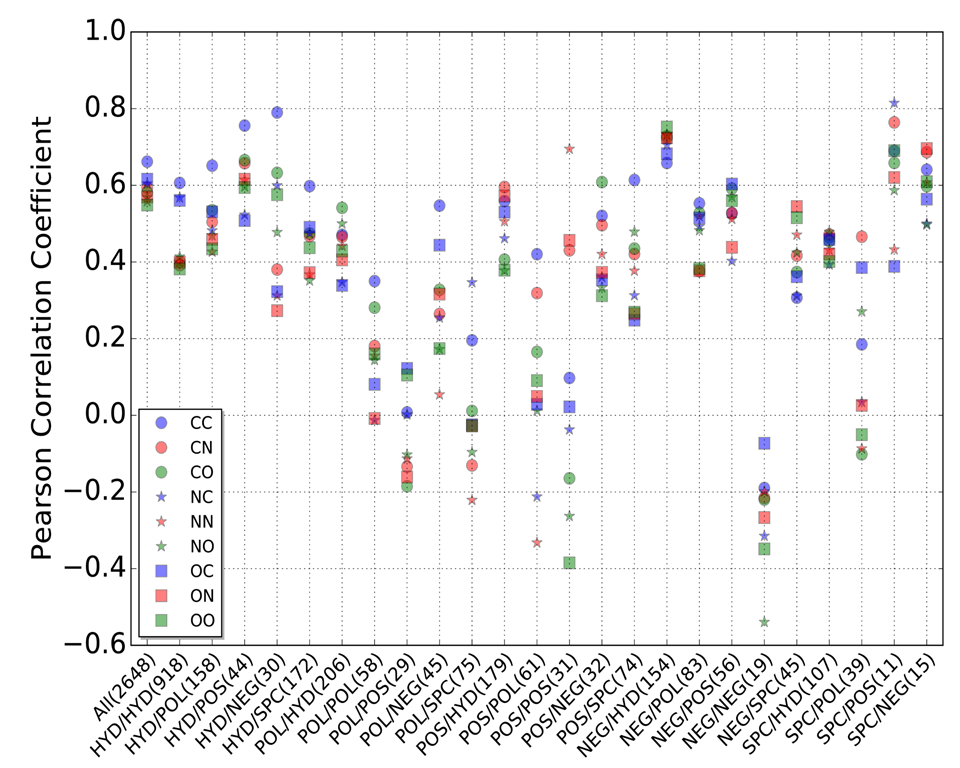}  
	\caption{Comparisons of the Pearson correlation coefficients obtained with 9 sets of ESPH features for the S2648 data set and its 24 subsets. The performance are the medians of 50 repeated runs.
		}.  
	\label{fig:Vperformance}
\end{figure}
To facilitate the discussion of different features, we denote topological features extracted from the atom set containing atoms of element type $\alpha$ in the rest of the protein and atoms of element type $\beta$ in the mutation site by $F_{\alpha\beta}$.
Typically, a more important feature has a higher  predictive power. Therefore, it is interesting to analyze  the predictive powers of individual interactive ESPH features (i.e., $F_{\rm CC}$, $F_{\rm CN}$, $F_{\rm CO}$, $F_{\rm NC}$, $F_{\rm NN}$, $F_{\rm NO}$, $F_{\rm OC}$, $F_{\rm ON}$, $F_{\rm OO}$). In this analysis, 
we consider  10-fold cross validations for the S2648 set and its subsets due to the small size of some subsets. Random forest regression with 3000 trees is used to reduce computation time. In each analysis, we use only one set of interactive ESPH features, such as $F_{\rm CC}$, to test the predictive power of hydrophobic features. The left column in  Figure \ref{fig:Vperformance} depicts our findings based on the whole data set of 2648 entries.  It is found that  features associated carbon atoms  in the mutation site, i.e., $F_{\alpha{\rm C}}$, give rise to some of the best predictions with Pearson correlation coefficients being higher than 0.65 (blue color ones). In fact, $F_{\rm CC}$ gives the best prediction, indicating the key importance of hydrophobic interactions to mutations.  Other features have  similar performances with  Pearson correlation coefficients ranging from  0.55 to 0.58.

We further analyze individual interactive ESPH feature performance with respect to different types of mutations.  The same classification of residue types is used as discussed in Section \ref{sec:PVMS}. We use HYD/POL to represent the situation in which a hydrophobic residue is mutated to a polar residue and similar notations are used  for other  situations. Our results are also presented  in Figure \ref{fig:Vperformance}.  Firstly, for 9 sets of mutation data that  involve hydrophobic residues, features that involve carbon atoms in the mutation site (i.e., $F_{\alpha{\rm C}}$) have a relatively high predictive power. Note that carbon atoms play a major role in hydrophobic interactions and changes in hydrophobic residues can be captured by the changes in Betti-0, Betti-1 and Betti-2 barcodes involving carbon atoms. In fact, other topological features do a good job in predicting hydrophobic residue involved mutations because this set of mutations leads to significant changes in topological invariants.  Secondly, all features that involve nitrogen atoms in the mutation site (i.e., $F_{\alpha{\rm N}}$) have a better predictive power for all positively charged residues (POS/POS). This occurs because three positively charged residues can be distinguished by their numbers of nitrogen atoms, which in turn, can be captured by Betti-0 barcodes   (i.e., $V ^{{\rm VC},DI,0}_{\gamma,\alpha,{\rm N}}$). Features constructed from oxygen atoms in the mutation site (i.e., $F_{\alpha{\rm O}}$) have the least prediction power for this data set. 
Thirdly, for mutations from one negatively charged residue to another negatively charged residue (i.e., NEG/NEG),  features constructed from nitrogen atoms in protein and oxygen atoms in the mutation site (i.e., $F_{\rm NO}$) have the worst predictive power. In fact, none of other topological features does a good job either. This poor performance might be due to negligible mutation induced changes in geometry, topology and structural stability. In this case, small changes in free energies are most likely caused by electrostatic redistribution, which is relatively insensitive  to the present topological description.  
Fourthly, all of the 9 types of features have a similar predictive power for the NEG/HYD data set.  Finally, small data size is hardly a pivotal factor in 10-fold cross validations, although all of the 7 lowest prediction data sets have  relatively small data sizes. Note that data sizes in all of the three best predictions (HYD/POS, HYD/NEG, SPC/POS) are fewer than 45 instances.   

\subsection{Performance on membrane proteins}

\begin{table}[ht]
\centering
\rowcolors{2}{gray!25}{white}
\begin{tabular}{lll}
\toprule
\rowcolor{gray!75}
Method & $R_P$ & RMSE\\
\midrule
T-MP-2$$ & \MemALLPCCMedian & \MemALLRMSEMedian\\
T-MP-1$$ & \MemTOPPCCMedian & \MemTOPRMSEMedian\\
Rosetta-MP         & 0.31 & - \\
Rosetta (High)$^a$ & 0.28 & - \\
FoldX              & 0.26 & 2.56 \\
PROVEAN            & 0.26 & 4.23 \\
Rosetta-MPddG      & 0.19 & - \\
Rosetta (low)$^b$  & 0.18 & - \\
SDM                & 0.09 & 2.40 \\
\bottomrule
\end{tabular}
\vspace{3mm}
\caption{Comparison of  Pearson correlation coefficients ($R_P$) and RMSEs (kcal/mol) of various methods for the M223 data set 
obtained from 5-fold cross validation. Except for the present results for T-MP-1 and T-MP-2, all other results are adopted from Kroncke \emph{et al}\cite{Kroncke:2016}. The results of Rosetta methods are obtained from Fig. S1 of Ref. \cite{Kroncke:2016} where RMSE is not given. The results of other methods are obtained from Table S1 of Ref. \cite{Kroncke:2016}. The results of the machine learning based methods are not listed since those servers are not trained on membrane protein data sets. Among the methods listed, only Rosetta methods have terms describing the membrane protein system. The results reported for T-MP methods are the median values of 50 repeated experiments. $^a$ High resolution. $^b$ Low resolution.  
}
\label{tab:MemMutationPerformance}
\end{table}

We also examine  performance of the proposed topological methods on a challenge problem identified by \cite{Kroncke:2016}. 
The proposed method is tested with 5-fold cross validations of a set of 223 mutation instances of membrane proteins in 7 protein families named  M223 data set   \cite{Kroncke:2016}. A comparison of  Pearson  correlation coefficients and  RMSEs over a number of methods  is shown in Table \ref{tab:MemMutationPerformance}. The machine learning based methods are not listed as they are trained on soluble protein data sets. As noted by Kroncke {\it et al}, there is no reliable method for the prediction of  membrane protein mutation impacts at present \cite{Kroncke:2016}. Nevertheless, our topology based approaches significantly outperform other existing physical or empirical methods. When auxiliary features are used together with topological features, a 5\% improvement in Pearson  correlation coefficient is found. Compared with Rosetta-MP, which achieves the best performance with terms designed for membrane proteins \cite{Kroncke:2016}, the present T-MP-2 has a 84\% higher Pearson correlation coefficient.  Nonetheless,  Kroncke {\it et al}'s statement about membrane protein mutation impact predictions still holds as the best Pearson correlation coefficient is only 0.57 and the best RMSE is over 1 kcal/mol. We therefore call for further methodology developments to improve membrane protein mutation impact predictions.

%
%

\section{Conclusion}
Contrary to geometry that dominates most biomolecular descriptions, topology is rarely implemented in quantitative  analysis of biomolecular science, due to its high level abstraction and dramatic reduction of biologic information. This article introduces element specific persistent homology to appropriately simplify biomolecular complexity while effectively retain essential biological information in protein mutation impact predictions. Extensive numerical experiments indicate that  element specific persistent homology offers some of the most efficient descriptions of protein mutation impacts that cannot be obtained by other conventional techniques.

\section*{Acknowledgments}

This work was supported in part by NSF Grants DMS-1160352 and IIS-1302285     and
MSU Center for Mathematical Molecular Biosciences Initiative.

\bibliographystyle{ieeetr}
\bibliography{refs}

\end{document}